# POSTURAL CONTROL DURING QUIET STANDING FOLLOWING CERVICAL MUSCULAR FATIGUE: EFFECTS OF CHANGES IN SENSORY INPUTS

Nicolas VUILLERME [1,2], Nicolas PINSAULT [3] and Jacques VAILLANT [3]

[1] Laboratoire de Modélisation des Activités Sportives, Université de Savoie, France.

[2] Laboratoire TIMC-IMAG, UMR UJF CNRS 5525, La Tronche, France

[3] École de Kinésithérapie du Centre Hospitalier Universitaire de Grenoble, France.

Address for correspondence:

Nicolas VUILLERME

Laboratoire TIMC-IMAG, UMR UJF CNRS 5525

Faculté de Médecine

38706 La Tronche cédex

France.

Tel: (33) (0) 4 76 63 74 86

Fax: (33) (0) 4 76 51 86 67

Email: nicolas.vuillerme@imag.fr

**Abstract**

The purpose of the present experiment was to investigate the effects of cervical muscular fatigue on postural control during quiet standing under different conditions of reliability and/or availability of somatosensory inputs from the plantar soles and the ankles and visual information. To this aim, 14 young healthy adults were asked to sway as little as possible in three sensory conditions (No vision, No vision-Foam support and Vision) executed in two conditions of No fatigue and Fatigue of the scapula elevator muscles. Centre of foot pressure (CoP) displacements were recorded using a force platform. Results showed that (1) the cervical muscular fatigue yielded increased CoP displacements in the absence of vision, (2) this effect was more accentuated when somatosensation was degraded by standing on a foam surface and (3) the availability of vision allowed the individuals to suppress this destabilising effect. On the whole, these findings not only stress the importance of intact cervical neuromuscular function on postural control during quiet standing, but also suggest a reweigthing of sensory cues in balance control following cervical muscular fatigue by increasing the reliance on the somatosensory inputs from the plantar soles and the ankles and visual information.

The maintenance of an upright stance is a complex task that requires the integration of visual, vestibular and somatosensory inputs from all over the body to assess the position and motion of the body in space and the ability to generate forces to control body position (e.g., [16]). The ability to maintain upright stance in a wide range of situations evidences that postural control system is adaptive to various task parameters (e.g., [18]) of either external or internal origin and can adjust postural responses by changing the relative weighting of sensory inputs (e.g., [10]). On the one hand, during quiet standing, external constraints involve, in part, the availability and integrity of sensory inputs from the environment. After a modification of the available sensory input, the individuals need to reconfigure their postural control system to adapt to the changing conditions, that is, to redefine the respective contribution of the different sources of sensory information for maintaining balance. For instance, withdrawing vision forces the individuals to reorganise the hierarchy of the sensory information because vestibular and proprioceptive information become the only sources of sensory information available for regulating posture, whereas introducing a foam to the supporting surface, altering the reliability to somatosensory afferents to postural control, yields greater reliance on visual and vestibular information (e.g., [24]). On the other hand, among the internal constraints postural control system has to cope with, the muscular fatigue represents an inevitable phenomenon of physical, professional or recreational activities known to alter the peripheral proprioceptive system, the central processing of proprioception but also the force generating capacity (e.g., [27]). Along these lines, numerous recent studies have reported a decreased ability to maintain balance in bipedal stance after fatiguing exercise (e.g., [5,14,28,29]). Interestingly, in most published works, muscular fatigue was induced by exercise of the lower extremities, whereas the effects of other muscle groups remained underexposed.

The purpose of the present experiment was to investigate the effects of cervical muscular fatigue on postural control during quiet standing. More specifically, scapula elevator muscles were chosen since many occupations involve postures that are characterised by sustained static or repetitive loading to the neck and shoulders, in particular to the Trapezius muscle (e.g., [3]). In the context of a multisensory control of balance, we examined these postural effects under different conditions of reliability and/or availability of somatosensory inputs from the plantar soles and the ankles and visual information. It was hypothesised that (1) the cervical muscular fatigue yielded a decreased postural control in the absence of vision, (2) this effect was more accentuated when somatosensory information at the plantar soles and the ankles was degraded and (3) the availability of vision allowed the individuals to suppress this destabilising effect.

Fourteen male university students from the Ecole de Kinésithérapie du Centre Hospitalier Universitaire de Grenoble (mean age = 22.6±3.0 years; mean body weight = 73.4±8.9 kg; mean height 178.4±6.1 cm) voluntarily participated in the experiment. They gave their informed consent to the experimental procedure as required by the Helsinki declaration (1964) and the local Ethics Committee. None of the subjects presented any history of neck trauma, neck pain, whiplash injury, any type of musculoskeletal problems, neurological disease or vestibular impairment. All subjects also had normal or corrected-to-normal vision.

Subjects stood barefoot on a force platform in a natural position (feet abducted at 30°, heels separated by 3 cm), their arms hanging loosely by their sides. The force platform (Equi+, model PF01), constituted of an aluminium plate (80 cm each side) laying on three uniaxial load cells, was used to measure the displacements of the centre of foot pressure (CoP). Signals from the force platform were sampled at 64 Hz, amplified and converted from analogue to digital form.

Subject's task was to sway as little as possible in three sensory conditions of No vision, No vision-Foam and Vision. In the No vision conditions, they were asked to close their eyes and to keep their gaze in a straight-ahead direction. In the No vision-Foam condition, a 2-cm thick foam support surface, altering the quality and/or quantity of somatosensory information at the plantar sole and the ankle (e.g., [24]), was placed under the subjects' feet. In the Vision condition, subjects stared at the intersection of a black cross (20 cm×25 cm) placed onto the white wall distant 1m in front of them, at the eyes level.

These three sensory conditions were executed the same day before (No fatigue condition) and after a designated fatiguing exercise for scapula elevator muscles (Fatigue condition). The scapula elevator muscular fatigue of both sides was induced until maximal exhaustion by instructing the subjects to perform a "dumbbell shrug trap exercise", commonly used in fitness training to specifically involve Levator scapulae and Trapezius superior muscles. Subjects picked up a dumbbell set and stood upright with their feet shoulder width apart and their shoulders back. With the arms straight at the sides, they were asked to shrug their shoulders up as high as possible while keeping their head up and their arms locked, as many times as possible following the beat of a metronome (40 beats/min). For each subject, the load used during this exercise was adjusted to 30% of their maximal workload determined in a preliminary experimental session by adding weight until they were able to perform a single repetition. Verbal encouragement was given to ensure that subjects worked maximally. The fatigue level was reached when subjects were no more able to complete the exercise. Immediately on the cessation of exercise, the subjective exertion level was assessed through the Borg CR-10 scale [2]. Subjects rated their perceived fatigue in the cervical muscles as almost "extremely strong" (mean Borg ratings of 9.2). To ensure that all balance measurements in the Fatigue condition were obtained in a real fatigued state, the fatiguing exercise took place beside the force platform, so that there was a short time-lag between the exercise-induced fatiguing activity and the balance measurements (less than 20 s) and the fatiguing exercise was repeated prior to each trial (e.g., [14,28,29]).

For each sensory condition (No vision, No vision-Foam and Vision) and each condition of No fatigue and Fatigue, subjects performed three 32-s trials, for a total of 18 trials. The order of presentation of the three sensory conditions was randomised over subjects.

Two dependent variables were used to describe subject's postural behaviour. (1) The mean medio-lateral (ML) and antero-posterior (AP) CoP positions (mm) represent the average position of CoP for the ML and AP axes, respectively. The mean ML CoP position was expressed relative to the mid-saggital line of the base of support, defined as the line perpendicular to the heel line passing through the geometric centre of base of support. The mean AP CoP position was measured from the heel line. Positive values for mean ML and AP CoP positions indicate a right and an anterior mean position of the COP relative to the mid-saggital line of the base of support and to the heel line, respectively. (2) The surface area ($mm^2$) covered by the trajectory of the CoP with a 90% confidence interval [25] is a measure of the CoP spatial variability. During quiet standing, increased values in CoP surface area indicate a decreased postural control, whereas decreased values express an increased postural control.

The means of the three trials performed in each of the six experimental conditions were used for statistical analyses. Two Fatigues (No fatigue versus Fatigue) × 3 Sensory conditions (No vision versus No vision-Foam versus Vision) analyses of variance (ANOVAs) with repeated measures on both factors were applied to the data. Post hoc analyses (Newman–Keuls) were used whenever necessary. Level of significance was set at 0.05.

Analysis of the mean position of the CoP along both the ML and AP axes did not show any significant interaction of Fatigue × Sensory condition ($F(2,26) = 0.26$, $P > 0.05$ and $F(2,26) = 0.57$, $P > 0.05$, for mean ML and AP CoP positions, Fig. 1A and B, respectively), nor significant main effects of Fatigue (F(1,13) = 0.00, P > 0.05 and F(1,13) = 0.16, P > 0.05, for mean ML and AP CoP positions, respectively) and Sensory condition (F(2,26) = 0.69, P > 0.05 and F(2,26) = 1.31, P > 0.05, for mean ML and AP CoP positions, respectively), ruling out the possibility the results observed through this study to be confounded by a possible effect of asymmetric [7] and leaning postures [21], respectively.

Analysis of the surface area covered by the trajectory of the CoP showed a significant interaction of Fatigue × Sensory condition (F(2,26) = 5.22, P < 0.05) (Fig. 2). The decomposition of this interaction into its simple main effects indicated that (1) the Fatigue condition yielded larger CoP surface area relative to the No fatigue condition in the No vision condition (P < 0.01), (2) this effect was more accentuated in the No Vision-Foam condition (P < 0.001), whereas (3) no significant difference between the two conditions of No fatigue and Fatigue was observed in the Vision condition (P > 0.05). The ANOVAs also confirmed main effects of Fatigue (F(1,13) = 31.42, P < 0.01) and Sensory condition (F(2,26) = 93.95, P < 0.001).

The purpose of the present experiment was to investigate the effects of cervical muscular fatigue on postural control during quiet standing under different sensory conditions. Somatosensory inputs from the plantar soles and the ankles were degraded by having subjects stand on a compliant surface (e.g., [24]), whereas visual information was either available or withdrawn.

In the absence of vision (No vision condition), results showed a decreased postural control following the cervical muscular fatigue, as assessed by a wider surface area covered by the CoP trajectory observed in the Fatigue than No Fatigue condition (Fig. 2, left part). These data are consistent with previous reports of increased CoP displacements following to neck muscle fatigue when vision was absent [9,22], hence confirming our first hypothesis. Although the exact mechanism inducing increased CoP displacements exercise is rather difficult to answer, it is possible that proprioception at the cervical joint was altered by the fatiguing exercise. Without direct measures of the cervical proprioceptive function, such a proposal is yet speculative. A study investigating whether the position sense at the neck is modified consecutive to muscular fatigue may provide more insight on this issue and is included in our immediate plans. At this point, however, alterations of proprioceptive functions at the ankle [6], knee [23], lumbar level [26], shoulder [1], or elbow [30] previously have been reported after a muscular fatiguing exercise. Along these lines, although lower limbs muscles are involved more directly on postural control during quiet standing (e.g., [31]), this result underlines the role of cervical proprioceptive input in postural control, in accordance with the assumption that this input operates through a "proprioceptive chain" linking the eye to the foot [20]. Indeed, when the reliance of cervical proprioceptive information is impaired by pathology, trauma or injury (e.g., [11,12,15,32]) or by experimental manipulations in normal subjects, such as neck muscles vibration (e.g., [8,13,20]), postural control has been shown to decrease.

In the context of the multisensory control of balance, when the availability or the reliability of input from a particular body location decreases, it is conceivable the central nervous system to increase the weighting of input from other locations that provide reliable information for maintaining stable posture (e.g., [10]). In our case, one way to compensate for cervical neuromuscular impairments consecutive to the fatiguing exercise could be to increase the reliance on alternative sensory systems such as somatosensation from the plantar soles or the ankles and vision.

On the one hand, while cervical muscular fatigue yielded a decreased postural control in the No vision condition (Fig. 2, left part), a greater destabilising effect was observed in the No vision-Foam condition, i.e., when somatosensory information was altered by the compliant support surface (Fig. 2, middle part). These results suggest an increased reliance on somatosensory inputs from the plantar soles and the ankles in the Fatigue relative to the No fatigue condition, confirming our second hypothesis. Interestingly, such a functional reweighting of proprioceptive sensory system controlling posture during quiet standing is not limited to neuromuscular impairments at the cervical region. Indeed, it has been recently demonstrated that low back pain lead to changes in postural control by refocusing proprioceptive sensitivity from the trunk to the ankles [4].

On the other hand, when vision was available (Vision condition), the increased CoP displacements observed following to the fatiguing exercise in the No vision conditions was cancelled out (Fig. 2, right part). This result suggests that the contribution of vision compensated for the effect of fatigue and allowed individuals to reach comparable levels of CoP displacements as those observed in the No fatigue condition, confirming recent result of Schieppati et al. [22] and our third hypothesis. The observation that vision had such a "restabilising" effect is not surprising given the major impact of visual input in maintaining upright stance (e.g., [19]). Again, such a sensory reweighting of visual information for controlling posture is not limited to cervical neuromuscular impairments. Indeed, the availability of vision has been shown to suppress postural differences observed between patients suffering from low back pain and age-matched healthy control subjects in absence of vision [17].

Altogether, these results suggest that the relative contributions of the somatosensory and visual systems in balance control differ between the two conditions of No fatigue and Fatigue. Precisely, an increased reliance on somatosensory inputs from the plantar soles and ankles and visual information was observed following cervical muscular fatigue. With regard to the hypothesis of an impairment of cervical neuromuscular function, these results are similar to recent findings in patients with cervical problems [12,15,32]. Indeed, compared to healthy control subjects, patients with neck trauma [12], chronic whiplash injury [15] or idiopathic cervical dystonia (ICD) [32] have been shown (1) to rely more on visual information for controlling upright stance and (2) to be more affected when the reliability of ankle proprioceptive information was reduced, either by standing on a compliant foam surface [32], providing inaccurate ankle sensory information by sway-referencing the support surface [12] or applying Achilles tendons vibrations [15] in absence of vision. In the study of W¨ober et al. [32], ICD patients also were tested before and after therapy with local injection of botulinum toxin type A (BTX-A). Six weeks after the therapy, ICD patients increased their postural control, a result that was suggested by the authors to stem from a reduction of abnormal proprioceptive inputs from the neck in patients. Interestingly, these post-treatment improvements were most marked during stance on foam with eyes closed and least during stance on firm surface with eyes open, suggesting that these two sensory conditions are probably the most and the least sensitive to abnormal proprioceptive input from the neck, respectively, as observed in the present experiment. These findings and ours thus argue in favour of a sensory reorganisation of postural control characterised by reweighting available sensory cues as a function of the neuromuscular constraints acting on the subject. They also suggest that, in conditions of neuromuscular impairments at the cervical region induced by muscular fatigue, injury, trauma or disease, the availability and integrity of somatosensory inputs from the plantar soles and the ankles, but especially visual information, is of great importance in the appropriate control of balance.

In conclusion, results of the present experiment showed that (1) the cervical muscular fatigue yielded increased CoP displacements in the absence of vision, (2) this effect was more accentuated when somatosensory information was disrupted by standing on a foam surface and (3) the availability of vision allowed the individuals to suppress this destabilising effect. On the whole, these findings not only stress the importance of intact cervical muscle function on postural control during quiet standing, but also suggest a reweigthing of sensory cues in balance control following to cervical muscular fatigue by increasing the reliance on the somatosensory inputs from the plantar soles and the ankles and visual information. Finally, we would like to mention that some subjects reported sensation of cervical pain at the end of the fatiguing exercise. Indeed, pain often develops following fatiguing muscle contractions. This sensation probably arises from firing of the groups III and IV afferents, that are sensitive to metabolites and inflammatory substances (e.g., potassium, lactic acid, bradykinin and arachidonic acid) accumulated within the muscle during activity to fatigue (e.g., [27]). There is thus a possibility that pain per se might affect postural control. Such a proposal is yet speculative and warrants additional investigations.

**Acknowledgements**

This paper was written while the first author was A.T.E.R. at Université de Savoie, France. The authors would like to thank subject volunteers and the anonymous reviewers for helpful comments and suggestions. Special thanks also are extended to Lucas B. for various contributions.

**Figure captions**

**Figure 1.** Mean and standard deviation of the mean position of the CoP along the mediol-lateral (A) and antero-posterior (B) axes for the two conditions of No fatigue and Fatigue and the three sensory conditions of No vision, No vision - Foam and Vision. The two conditions of No fatigue and Fatigue are presented with different symbols: No fatigue (*black diamond*) and Fatigue (*white circle*). Note that the mean ML and AP CoP positions were not affected by the experimental conditions.

**Figure 2.** Mean and standard deviation of surface area covered by the trajectory of the CoP displacements for the two conditions of No fatigue and Fatigue and the three sensory conditions of No vision, No vision - Foam and Vision. The two conditions of No fatigue and Fatigue are presented with different symbols: No fatigue (*black diamond*) and Fatigue (*white circle*). The significant *P*-values for comparisons between No fatigue and Fatigue conditions also are reported (**: $P<0.01$; ***: $P<0.001$).

**Figure 1**

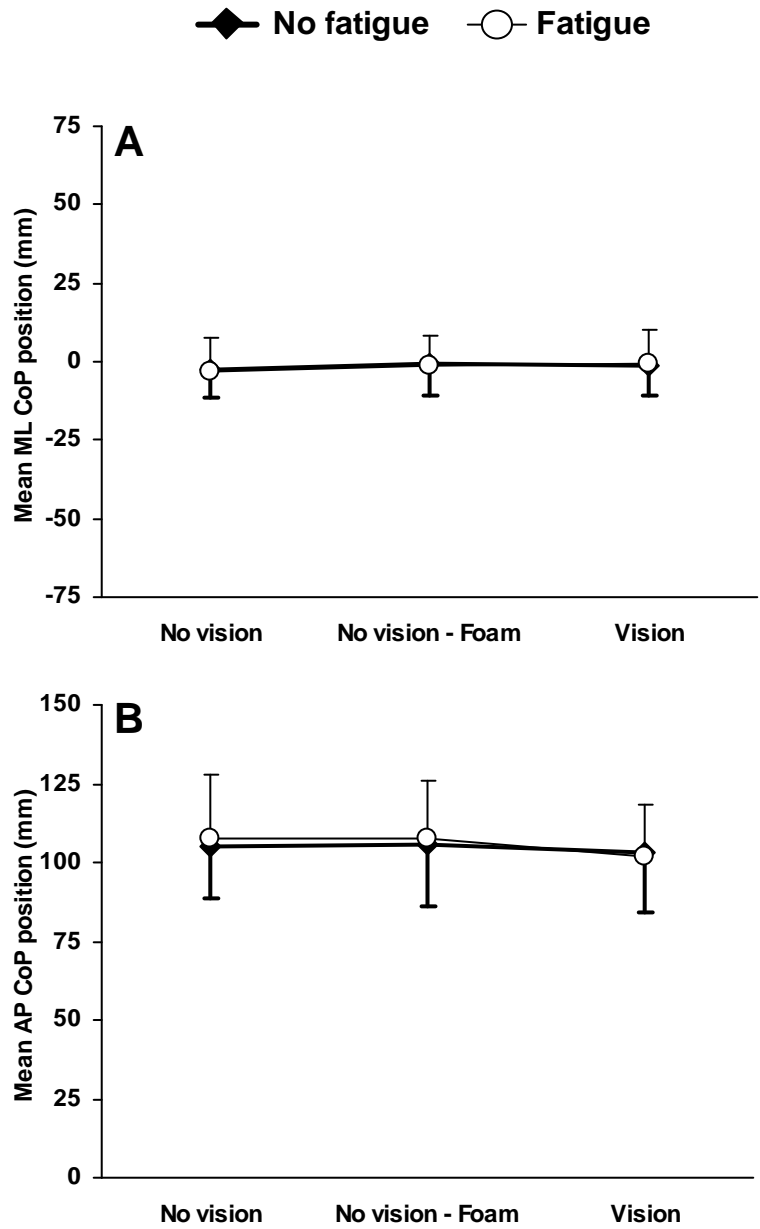

**Figure 2**

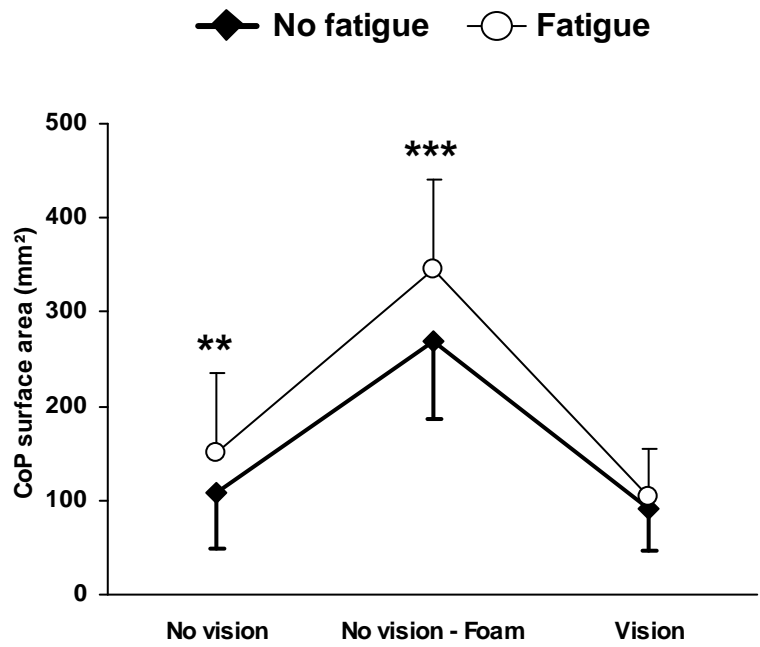